\documentclass[oldversion]{aa}
\usepackage{epsfig}
\usepackage{natbib}
\newcommand{\lya}{\mbox{Ly$\alpha$}}

\titlerunning{Spatially-resolved LBG at z$\ge 3$}
\authorrunning{Nesvadba et al.}
\begin{document}

\title{Integral-field spectroscopy of a Lyman-Break Galaxy
  at z$=$3.2: evidence for merging}

\author{N.~P.~H. Nesvadba\inst{1,2}, M.~D.~Lehnert\inst{1},
  R.~I.~Davies\inst{3}, A.~Verma\inst{4}, F.~Eisenhauer\inst{3}} 

\institute{Observatoire de Paris, CNRS, Universite Denis Diderot; 5,
  Place Jules Janssen, 92190 Meudon, France 
\and 
Marie-Curie Fellow
\and
Max-Planck-Institut f\"ur Extraterrestrische Physik, Garching bei M\"unchen,
  Germany
\and
University of Oxford, Subdepartment of Astrophysics, Denys Wilkinson Building,
Keble Road, Oxford, UK.
} 

\date{Recieved / Accepted}
\abstract{{}We present spatially-resolved, rest-frame optical
spectroscopy of a z$\sim 3$ Lyman-break galaxy (LBG), Q0347-383~C5,
obtained with SINFONI on the VLT. This galaxy, among the $\sim 10$\% brightest
LBGs, is only the second z$\sim 3$ LBG observed with an
integral-field spectrograph. It was first described by Pettini et
al. (2001), who obtained WFPC2 F702W imaging and longslit spectroscopy
in the K-band. We find that the emission line morphology is dominated
by two unresolved blobs at a projected distance of $\sim 5$ kpc with a
velocity offset of $\sim 33$ km s$^{-1}$. Velocity dispersions suggest
that each blob has a mass of $\sim 10^{10}$ M$_{\odot}$. Unlike
Pettini et al. (2001), our spectra are deep enough to detect H$\beta$,
and we derive star-formation rates of $\sim 20-40$ M$_{\odot}$
yr$^{-1}$, and use the H$\beta$/[OIII] ratio to crudely estimate an
oxygen abundance $12+[O/H]=7.9-8.6$, which is in the range typically
observed for LBGs.  We compare the properties of Q0347-383~C5 with
what is found for other LBGs, including the gravitationally lensed
``arc$+$core'' galaxy (Nesvadba et al. 2006), and discuss possible
scenarios for the nature of the source, namely disk rotation, a
starburst-driven wind, disk fragmentation, and merging of two LBGs. We
favor the merging interpretation for bright, extended LBGs like
Q0347-383~C5, in broad agreement with predicted merger rates from
hierarchical models.}

\keywords{cosmology: observations --- galaxies: evolution ---
galaxies: kinematics and dynamics --- infrared: galaxies} 

\maketitle

\section{Introduction}
\label{introduction}
Our understanding of high-redshift galaxies is growing at a rapid pace. By
combining ground and space-based observations, all of the electromagnetic
spectrum from the X-rays to the radio is now being used to select and study
galaxies in the early universe.  Among the ever-growing number of
high-redshift galaxy populations, Lyman Break Galaxies (LBGs) play a dominant
role, from an astrophysical and from an observational point of view. Recently,
\citet{reddy07} argued that blue, star-forming galaxies like LBGs may have had
the largest contribution to the bolometric energy output of galaxies at
redshifts z$\sim 2-3$, emphasizing their importance for our understanding of
the cosmic star-formation history.  Observationally, more than 1000 LBGs have
spectroscopic redshifts at z$\sim 3$ alone \citep[e.g.,][]{steidel03}, and
their ensemble properties, like their spatial clustering, luminosity function,
and average rest-frame UV spectral properties are known at unprecedented
precision \citep[e.g.,][]{adelberger05,shapley01,papovich01,shapley03}. Very
recently, \citet{law07} presented adaptive-optics assisted integral-field
spectroscopy of the $z= 3.3$ LBG DSF2237a-C2 using OSIRIS on the Keck
telescope. 

The picture that emerges from these observations is far from 
simple. \citet{law07} point out that at resolutions of $\sim 1$ kpc reached
with adaptive optics, UV-selected z$\sim 2-3$ galaxies have irregular
kinematics, which are likely not dominated by large-scale gravitational
motion, but perhaps are more related to merging or
gas-cooling. \citet{genzel06,nmfs06} however argue that at least a subsample
of blue, star-forming galaxies at somewhat lower redshifts, z$\sim 2$, may
show the signs of large, spatially-extended, rotating disks. Distinguishing
between the two scenarios is difficult, due to the low spatial resolution
of the data relative to the size of the targets \citep[see
also][]{kronberger07}. 

Here we present a study of one of the first z$\sim 3$ LBGs described
in the literature, Q0347-383~C5, which was initially described by
\citet{steidel96b}.  With ${\cal R}=$23.82 mag, Q0347-383~C5 is within
the tail of the $\sim 10$\% brightest LBGs.  WFPC2 imaging shows a
relatively complex morphology, extending over $\sim 1$\arcsec down
to the faint surface brightness detection limit of the image (18000s
through the F702W filter) and a half light radius about a few tenths of
an arc second \citet{pettini01}.  Extents this large are not rare among
the bright LBG population \citep{conselice03} and its half-light radius
is also rather typical \citep{ferguson04}.  \citet{pettini01} obtained
longslit spectroscopy in the K-band for a small sample of z$\sim 3$
LBGs, including Q0347-383~C5, which was one out of two of their sources
with spatially-extended spectra and a velocity gradient of $\sim 70$
km s$^{-1}$ in the [OIII]$\lambda\lambda$4959,5007 emission line. They
placed a 3$\sigma$ limit on H$\beta$ of $F(H\beta) < 1.7 \times 10^{-17}$
erg s$^{-1}$ cm$^{-2}$.

Using the near-infrared spectrograph SINFONI on the VLT, we obtained deep,
spatially-resolved spectroscopy of the [OIII]$\lambda\lambda$4959,5007,
and H$\beta$ emission line. This is only the second unlensed LBG with
integral-field spectroscopy in the literature with such observations
\citep[][discuss the z=3.2 LBG DSF2238a-C2, for which they obtained
rest-frame optical integral-field spectroscopy using OSIRIS on the Keck
with a laser guide star.]{law07} Given the small number of LBGs with
integral-field spectroscopy, and the large observational expense of such
observations, even the study of a single object is already a significant
step forward. Such observations overcome many uncertainties related
to longslit spectroscopy, such as slit-losses, and allow us to trace
the emission line morphology and kinematics across the two-dimensional
surface of the target. This makes them particularly suited to disentangle
the often complex emission line morphology of high-redshift galaxies.

Throughout the paper we adopt a flat H$_0 =$70 km s$^{-1}$ Mpc$^{-3}$
concordance cosmology with $\Omega_{\Lambda} = 0.7$ and $\Omega_{M} = 0.3$. In
this cosmology, D$_L=$ 28 Gpc and D$_A =$ 1.5 Gpc at z=3.23. The size scale is
7.5 kpc arcsec$^{-1}$. The age of the universe at this redshift and
cosmological model is 1.9 Gyr.

\section{Observations and data reduction}
\label{sec:observations}
We obtained deep, seeing-limited K-band spectroscopy of the z$=$3.2
LBG Q0347-383~C5, using the integral-field spectrograph SINFONI
\citep{eisenhauer03,bonnet04} on the VLT in December 2004 under excellent
and stable conditions. SINFONI is a medium-resolution, image-slicing
integral-field spectrograph, with 8\arcsec$\times$ 8\arcsec\ field of
view at a 0.125\arcsec$\times$ 0.125\arcsec\ pixel scale and spectral
resolving power $R \sim 4000$ in the K band. The total observing time
was 14400s with individual exposure times of 600s. This corresponds to
the on-source observing time, since we adopted a dither pattern where
the galaxy was constantly within the field of view.

We used the IRAF \citep{tody93} standard tools for the reduction
of longslit-spectra, modified to meet the special requirements of
integral-field spectroscopy, and complemented by a dedicated set of IDL
routines. Data are dark-frame subtracted and flat-fielded. The position
of each slitlet is measured from a set of standard SINFONI calibration
data, measuring the position of an artificial point source. Rectification
along the spectral dimension and wavelength calibration are done before
night sky subtraction to account for some spectral flexure between the
frames. Curvature is measured and removed using an arc lamp, before
shifting the spectra to an absolute (vacuum) wavelength scale with
reference to the OH lines in the data. To account for variations in
the night sky emission, we normalize the sky frame to the average of
the object frame separately for each wavelength before sky subtraction,
correcting for residuals of the background subtraction and uncertainties
in the flux calibration by subsequently subtracting the (empty sky)
background separately from each wavelength plane.

The three-dimensional data are then reconstructed and spatially aligned
using the telescope offsets as recorded in the header within the same
sequence of 6 dithered exposures (about one hour of exposure),
and by cross-correlating the line images from the combined data in each
sequence, to eliminate relative offsets between different
sequences.  Telluric correction is applied to each individual cube
before the cube combination. Flux scales are obtained from standard
star observations taken every hour at similar position and air mass as
the source.

We also used the standard star to carefully monitor the seeing during
observations, and we find an effective seeing in the combined cube of
FWHM 0.55\arcsec$\pm0.05$\arcsec$\times$0.49\arcsec$\pm$0.04\arcsec. The
spectral resolution was measured from night-sky lines and is FWHM$=$103 km
s$^{-1}$ at the wavelength of [OIII]$\lambda$5007.

\subsection{Additional data sets and alignment}
We also obtained WFPC2 F702W imaging of Q0347-383~C5 from the HST archive,
which was originally presented by \citet{pettini01}. We used the OTFC
calibrated data sets, and followed the standard DRIZZLE procedures given
in, e.g., \citet{drizzlemanual}, to remove cosmic rays and to align and
combine the individual frames.

It is difficult to accurately align the SINFONI and WFPC2 data at
sub-arcsecond precision, because of the small field of view of SINFONI
of only 8\arcsec$\times$8\arcsec\ and the small source size, which is
about similar to the uncertainty in the absolute astrometry of both the
VLT and WFPC2 ($\sim 1\arcsec$). Moreover, the morphologies in the WFPC2
continuum image and SINFONI line image are very different.

We therefore base the alignment on astrophysical arguments. Overall,
the emission line regions will roughly align with the continuum, as
typically observed in blue, star-forming galaxies at redshifts z$\sim
2$ \citep{nmfs06,law07,nesvadba07}.  The only high-redshift galaxies,
where the line and continuum emission do not seem to align well, are
radio galaxies in which several $10^{10}$ M$_{\odot}$ of ionized
gas extend over radii of $\sim 20-30$ kpc, which appear to be entrained
and ionized by feedback from the powerful AGN \citep{nesvadba07}. But
those can hardly be good analogs to Q0347-383~C5.

If we align the unresolved, bright knot in the southern part of the
source with one of the unresolved knots in the [OIII]$\lambda$5007
emission line image, the line and continuum emission are overall well
aligned. Astrophysically, this particular choice relies on the assumption
that the strong line emission and UV continuum originate from the same
region, which is a reasonable assumption for both star-forming regions,
and AGN.  The spatial extent of the line and continuum emission are
well matched, which may serve as a heuristic justification of the
method, and this choice does not have a strong impact on our overall
interpretation. We note that the distance between the two knots in the
line image does not correspond to the distance between the faint source
towards the north, which is marginally detected in the WFPC2 image,
and any part of Q0347-383~C5.

\section{Spatially-resolved integral-field spectroscopy of a Lyman-Break
  Galaxy at z=3.2} 
\label{sec:results}
\subsection{Continuum and emission line morphology}
\begin{figure}
\centering
\epsfig{figure=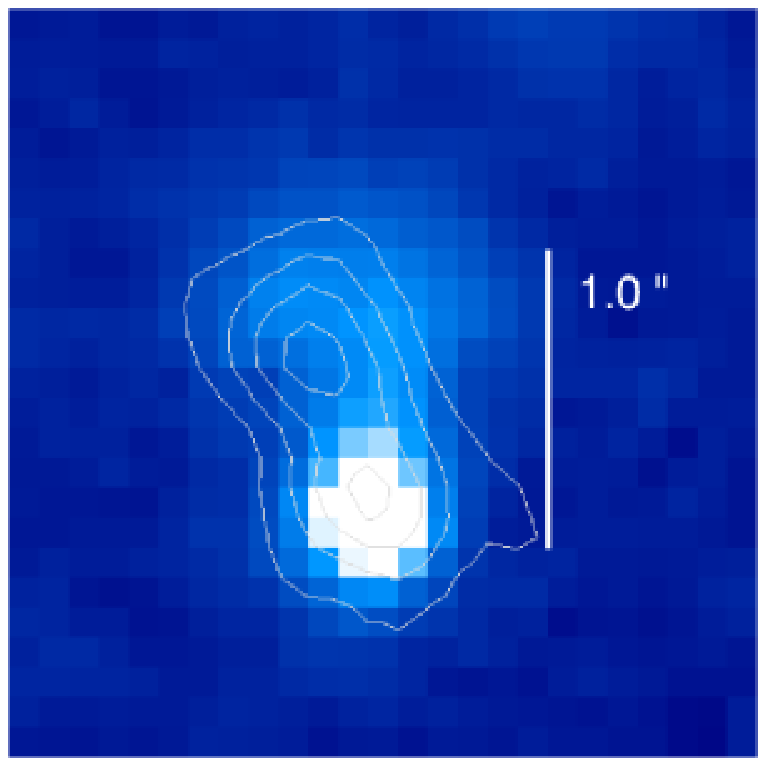,width=0.23\textwidth}
\epsfig{figure=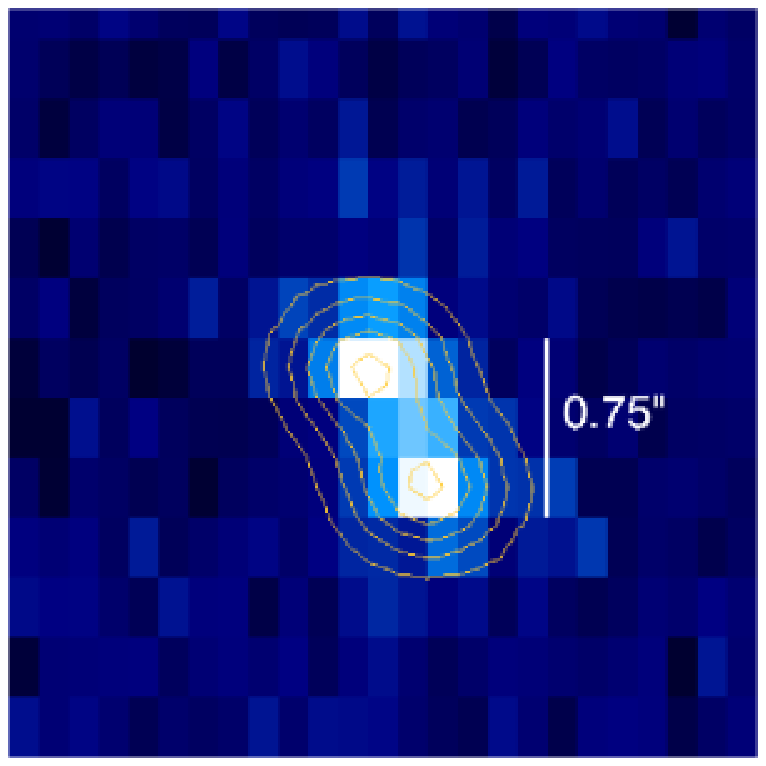,width=0.23\textwidth}
\caption{{\it left:} [OIII]$\lambda$5007 rest-frame UV morphology observed
with WFPC2 through the F702W filter first presented by
\citet{pettini01}. The contours indicate the [OIII]$\lambda$5007 emission
line morphology extracted from our SINFONI data cube.
{\it right:} [OIII]$\lambda$5007 emission line morphology of
Q0347-383~C5. 
For comparison, contours indicate the morphology of two artificial
point-sources at the position of each knot in the [OIII] line image convolved 
with two-dimensional Gaussian distributions with widths corresponding to the
spatial resolution of the data.}
\label{fig:o3im}
\end{figure}

Q0347-383~C5 is one of the largest z$\sim 3$ LBGs, in particular
it is large enough to be spatially resolved with seeing-limited
observations. Fig.~\ref{fig:o3im} shows the F702W morphology obtained with
the HST. At z=3.23, the F702W bandpass corresponds to rest-frame
wavelengths of $\sim 1470-1800$\AA. The brightest emission comes
from a compact, marginally resolved knot, which has an intrinsic size
of 0.13\arcsec$\times$0.18\arcsec\ (0.97kpc $\times$ 1.49 kpc for our
cosmology). We obtained this estimate from assuming that the observed full
width at half maximum of the image is a quadratic sum of the intrinsic
FWHM of the source and the point spread function, 0.1\arcsec$\times$
0.1\arcsec, which we measured from a nearby star on the same chip. The
knot is separated by $\sim 0.6$\arcsec\ (4.5 kpc) from a more diffuse,
elongated object.  About 1.6\arcsec (12 kpc) to the north-west from the
brightest knot is another diffuse source, which is marginally detected
in the WFPC2 image.

We show the [OIII]$\lambda$5007 emission line morphology of Q0347-383~C5
in Figure~\ref{fig:o3im}. The image includes wavelengths $\pm 1$ FWHM
around the peak integrated emission.
 The emission is clearly spatially extended over an
area of $\sim 0.6$\arcsec$\times1.3$\arcsec, corresponding to $\sim 4.5$
kpc$\times9.8$ kpc. We identify two separated, unresolved line emitters at
a projected distance $d_{proj} \sim 0.7$\arcsec\ or 5.3 kpc. Each of the
knots is spatially unresolved (see right panel of Fig.~\ref{fig:o3im}
which shows the line distribution compared to a point source), and
we place upper limits on their size from the size of the seeing disk,
finding FWHM of $< 2.7 \times 2.3$ kpc in right ascension and declination,
respectively. We do not detect any line emission from the diffuse,
faint continuum source to the north west.

\subsection{Integrated spectrum}
\begin{figure}
\centering
\epsfig{figure=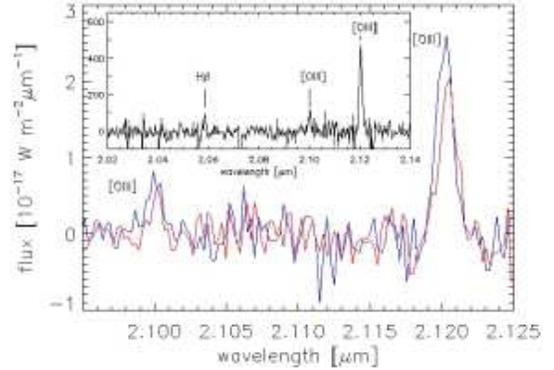,width=0.3\textwidth,angle=90}
\caption{Individual spectra of knot $A$ and $B$ are shown as red and blue
  curve, respectively. Both components have very similar spectral properties,
  and a relative velocity shift of $33\pm 10$ km s$^{-1}$. The inset shows the
  integrated spectrum of Q0347-383~C5 extracted from our
  SINFONI data cube. H$\beta$ and the [OIII]$\lambda\lambda$4959,5007 doublet
  are clearly detected.}
\label{fig:spectrum}
\end{figure}

The integrated spectrum of Q0347-383~C5 is shown in  the inset of
Fig.~\ref{fig:spectrum}. The spectrum was integrated by summing
over all spatial pixels in which the [OIII]$\lambda$5007 emission
exceeds 3$\sigma$. The redshift is z=3.2347$\pm$0.0007. The
[OIII]$\lambda\lambda$4959,5007 doublet is very prominent, and
detected at 5$\sigma$ and 12$\sigma$, respectively. Our results on
the redshift, FWHM, and line flux of [OIII] agree with those of
\citet{pettini01}, but unlike \citeauthor{pettini01}, we also detect
H$\beta$ with a flux of F$_{H\beta} = 8.8\pm 1.8 \times 10^{-18}$
erg s$^{-1}$ cm$^{-2}$. This flux is consistent with the 3$\sigma$
upper limit given by \citet{pettini01}. We find a [OIII]$/$H$\beta$
flux ratio of [OIII]$_{5007}$/$H\beta= 7.2\pm $1.5. Line widths
are FWHM$_{5007}$=180$\pm$9 km s$^{-1}$ for [OIII]$\lambda$5007, and
FWHM$_{H\beta}$= 69$\pm$11 km s$^{-1}$ for H$\beta$, respectively. The
lower H$\beta$ line width may be due to its unlucky wavelength with
respect to the telluric absorption.

We use the integrated spectrum to give a rough R$_{23}$-like metalicity
estimate for Q0347-383 C5. We did not measure the
[OII]$\lambda\lambda3726,3729$ doublet, therefore, we use the correlation of
[OII]$\lambda$3727/[OIII]$\lambda$5007 with [OIII]$\lambda$5007/H$\beta$ given
by \citet{kobulnicky99} for low-metalicity galaxies to estimate the most
likely [OII]$\lambda$3727 flux. With the measured uncertainties and the
$\log([OII]/H\beta) = 0.4$ suggested by the \citeauthor{kobulnicky99}
correlation, R$_{23}=1.05$.
If instead we only use the measured [OIII] and H$\beta$ values, and
  neglect any contribution from [OII], we find R$_{23,OIII} = 0.95$. This
corresponds to a highly conservative, but probably very loose lower bound.
Including the $1\sigma$ uncertainties of
our flux measurements, this corresponds to R$_{23}> 0.86$, or a metalicity
between 8.6 and 7.9.

This estimate may appear relatively uncertain, but we emphasize that this is
the case for any metalicity estimate of high-redshift galaxies from emission
lines. Even the sample of \citet{pettini01}, which had measured
[OII]$\lambda3727$, [OIII]$\lambda\lambda$4959,5007, and H$\beta$ fluxes,
could not be corrected for extinction, which will introduce considerable
uncertainties. With these caveats in mind, Q0347-383~C5 has an oxygen
abundance similar to those of the subsample of \citet{pettini01} with R$_{23}$
measured. Comparing with the solar oxygen abundance estimate of
\citet{allende01}, $[O/H] = 8.69\pm 0.05$, we find that Q0347-383~C5 has
a mildly subsolar metalicity ranging $[M/H]$$\sim$$-$0.1 -- $-$0.7dex.

\subsection{Spatially-resolved kinematics and properties of individual knots}
\label{ssec:kinematicsofknots}
\begin{figure}
\centering
\epsfig{figure=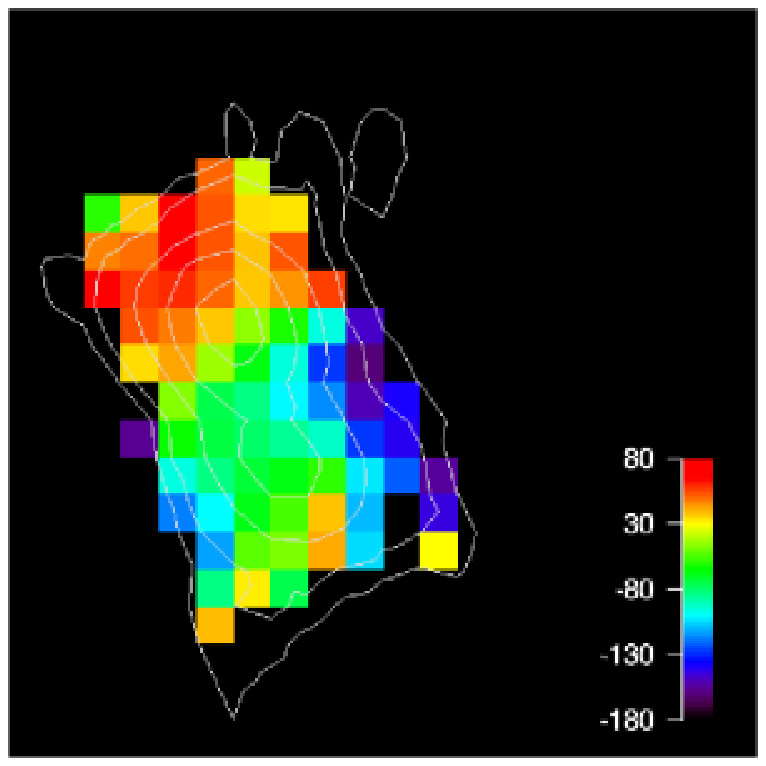,width=0.23\textwidth}
\epsfig{figure=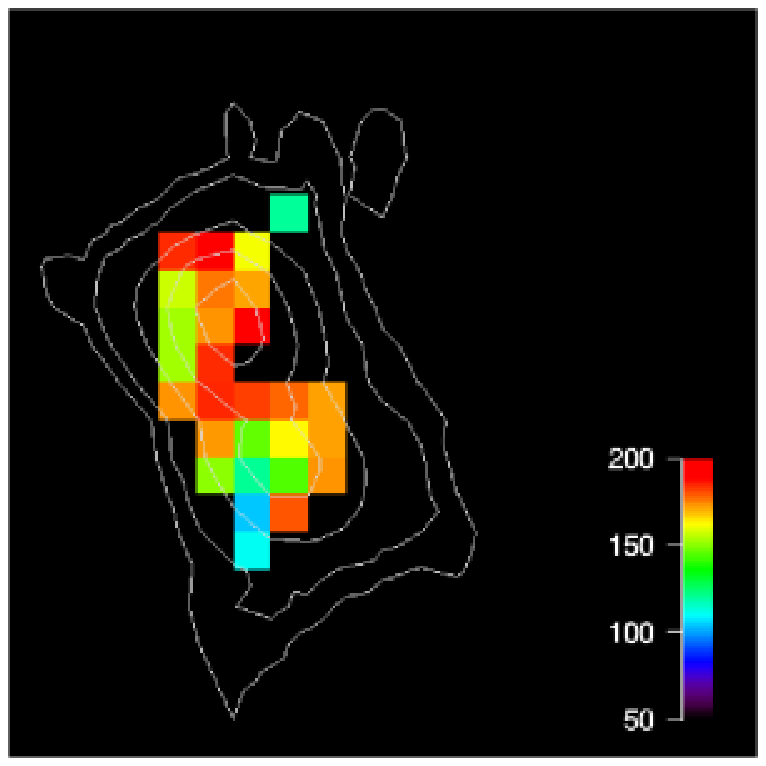,width=0.23\textwidth}
\caption{The velocity and width map of Q0347-383~C5 are shown in the left
and right panel, respectively. The velocity does not vary smoothly across the
source, but abruptly between the two knots. Line widths are uniform across the
source. Color bars indicate the velocities and FWHM line widths in km s$^{-1}$
in the left and right panel, respectively.} 
\label{fig:maps}
\end{figure}

To map the velocity and dispersion fields, we extracted the
[OIII]$\lambda$5007 emission lines from 3$\times$3 pixel apertures
(0.375\arcsec$\times$0.375\arcsec), which is below the seeing disk. In
addition, we smoothed the spectra over 3 pixels along the spectral axis. The
result is shown in Fig.~\ref{fig:maps}. Typical uncertainties are $\sim 10-15$
km s$^{-1}$ in both maps. Velocities between the two components are
significantly different, and do not vary smoothly, but change abruptly near
the mid point between the knots.
Velocities in each individual knot are very uniform across the full
two-dimensional surface. The [OIII] morphology further suggests that most of
the line emission comes from the two knots, with a negligible contribution
from more diffuse, extended areas. This correspondence between the
two-dimensional morphology and velocity map suggests that Q0347-383~C5 is
composed of two individual components, each dominated by its internal
kinematics. FWHM line widths are relatively uniform, and are
marginally lower in the northern component. Towards north-west, lines are
not resolved spectrally. 

We extracted spectra from each individual knot, finding that both
have very similar properties (Fig.~\ref{fig:spectrum}). For knot $A$
and $B$, we find line widths of FWHM$_A= 207\pm 7$ km s$^{-1}$
and FWHM$_B= 181\pm 3$ km s$^{-1}$, respectively, and a velocity
offset between the two integrated spectra of $\Delta v = 33\pm
10$ km s$^{-1}$. We observe a similar offset in both lines of the
[OIII]$\lambda\lambda$4959,5007 doublet. H$\beta$ is detected in both
components, with fluxes of F$_A$(H$\beta$) = $6.6\pm 1.8\times 10^{-21}$
W m$^{-2}$ and  F$_B$(H$\beta$) = $5.7\pm 1.8\times 10^{-21}$ W m$^{-2}$
in components $A$ and $B$, respectively. Oxygen ratios in the two knots
are similar within the (relatively large) uncertainties, suggesting
subsolar metallicities in both components.

We use the measured velocity dispersions, corrected for the instrumental
resolution, of each knot to give a rough estimate of the mass of each
component, assuming that each knot has a King profile and setting
$M=c\ \sigma^2\ R/G$ with dynamical mass, $M$, velocity dispersion,
$\sigma$, radius, $R$, and gravitational constant, $G$. $c$ is a
correction factor, with $c\sim 4-6$ for early-type galaxies on the
fundamental plane, depending on the ratio between tidal radius and core
radius \citep{bender92}. For simplicity, and since we are not able
to constrain this ratio from our observations, we assume $c=5$. The
additional uncertainty does not dominate the overall error budget of our
mass estimate. The FWHMs given in \S\ref{ssec:kinematicsofknots}
correspond to velocity dispersions in the two knots of $88\pm 3$ km
s$^{-1}$ and $77\pm 3$ km s$^{-1}$ (\S\ref{sec:results}), respectively,
and we use the upper limits on the size of each clump, $HWHM \le 1.4$
kpc. Thus, we find $M_A \la 9\times 10^{9} M_{\odot}$ and $M_B \la
1.2\times 10^{10} M_{\odot}$ for component $A$ and $B$, respectively. Note
that \citet{pettini01} used the same method to estimate the average
dynamical mass of LBGs, for an average radius of 2.5 kpc.

If alternatively, we assume that the velocity dispersion in each knot is
dominated by disk rotation, then we can set $M=v^ 2\ R/G,$ with
circular (and deprojected) velocity $v= f_c\times v_{obs}$, corresponding to
the observed velocity, $v_{obs}$, corrected by a factor $f_c = 1.7$, for
an average inclination and the apparent flattening of the rotation curve due
to the seeing \citep{rix97}. We find $M_A = 5\times 10^9
M_{\odot}$ and $M_B = 7\times 10^{9} M_{\odot}$, respectively.

Having detected H$\beta$, we can estimate star-formation rates in each
knot, following \citet{kennicutt98} and adopting a Balmer decrement
of H$\alpha$/H$\beta$=2.86. For a Salpeter IMF and mass range of $1
-100$ M$_{\odot}$, this corresponds to a conversion of star-formation
rate, SFR, to emission line luminosity, ${\cal L}_{H\beta}$, of SFR
[M$_{\odot}$ yr$^{-1}$] $= {\cal L}_{H\beta}$ [$3.7\times 10^{34}$
W]. For components $A$ and $B$, respectively, we find star formation
rates SFR$_A=14 M_{\odot}$ yr$^{-1}$ and SFR$_B=12 M_{\odot}$ yr$^{-1}$.
These values were derived with the assumption that extinction
in Q0347-383~C5 is negligible and therefore our estimates correspond
to lower limits. \citet{shapley01} found for their LBG sample E(B-V) =
0.2-0.4, indicating that intrinsic fluxes may be factors 1.6-3 higher
(for a galactic extinction law). 
We therefore do not expect that extinction corrected rates will greatly
 exceed SFR$_A\sim 19-35$ M$_{\odot}$ yr$^{-1}$ and SFR$_B\sim 22-42
M_{\odot}$ yr$^{-1}$ for components A and B, respectively. Using the observed
G$-$R color of Q0347-383~C5, G$-$R = 0.65 mag, and for a constant starburst
with an 
age of a few $10^7$ years, we expect extinctions that are even lower,
E(B-V)$<<$0.1. 

\section{Q0347-383 C5 -- Formation, outflow, or rotation?}
\label{sec:discussion}
Integral-field spectroscopy \citep[e.g.][]{law07,nmfs06} and longslit
spectroscopy \citep{erb03} revealed that galaxies at z$\sim2-3$ often have
complex kinematics that are not easily interpreted. While \citet{law07}
highlight the small observed ratios between velocity gradient and
total dispersion, v/$\sigma$, of many of the most extended sources,
and suggest that this may be due to processes more complex than simple
self-gravitating disks, \citet{nmfs06,genzel06} argue for rotation of
large-scale disks in at least a subset, the best observed subset, of
their sample. Evidence for rotation on sub-kpc scales, has been found
by \citet{nesvadba06b} in a gravitationally lensed, strongly magnified
arc at z=3.24. Alternatively, the gas kinematics may be influenced by
feedback from star-formation or AGN. We will in the following discuss
these possible interpretations.

\subsection{The wind scenario}
Many LBGs show blue asymmetries in the profiles of their rest-frame
optical emission lines \citep{nesvadba07}. Generally, in actively
star-forming galaxies, such asymmetries are interpreted as a sign of
starburst-driven outflows of ionized gas \citep{lehnert96}. Similar
outflows are expected in galaxies with star-formation densities $> 0.1
M_{\odot}$ yr$^{-1}$ kpc$^{-2}$ \citep{heckman03}, and most LBGs
easily surpass this limit. The star-formation intensities in
Q0347-383~C5 are above this limit ($>$2 M$_{\odot}$ yr$^{-1}$
kpc$^{-2}$, corresponding to the observed HWHM to approximate the
radius, and the measured star-formation rate, SFR$_B=12 M_{\odot}$
yr$^{-1}$, in component $B$).

However, the [OIII]$\lambda$5007 line wings do not show
well-pronounced blue wings, and the [OIII] emission line morphology in
Fig.~\ref{fig:o3im} does not suggest that the overall line emission is
dominated by a wind. The line emission is concentrated in two knots
that are each spatially unresolved. In contrast, the morphologies of
starburst-driven winds at low redshift \citep[e.g.,][]{lehnert96}
typically resemble edge-brightened bubbles in the line emission that
have ``broken out'' of the confinement provided by the ambient ISM.

Moreover, starburst-driven winds have typically low surface brightness
at low redshift \citep{lehnert96} and even in massive starbursts at
high redshift in spatially-resolved data sets. For the submillimeter
selected z=2.6 galaxy SMMJ14011+0252 with a star-formation rate of a
few $\times 100$ M$_{\odot}$ yr$^{-1}$, \citet{nesvadba07} find that
starburst-driven winds do overall not dramatically alter the observed,
large-scale kinematics of the galaxy, but have a measurable influence
on the line profiles of the optical emission lines.

Comparison with the rest-frame UV absorption line spectrum yields
similar conclusions. \citet{pettini01} give redshifts for \lya\ and the
interstellar absorption lines for Q0347-383~C5 of z$(Ly\alpha)=3.244$ and
z(abs)$ = 3.236$, respectively. While redshifted \lya\ and interstellar
lines relative to the rest-frame optical emission lines are commonly
interpreted as evidence for outflows of neutral and ionized gas, in
Q0347-383~C5, both Ly$\alpha$ and rest-frame UV absorption lines have
redshifts relative to the rest-frame optical emission line gas. This may
indicate more complex kinematics of the neutral and ionized material,
perhaps related to infalling gas, or to a variety of physical processes
affecting the kinematics of different components of the gas in LBGs.

\subsection{The rotating disk scenario}
Empirically, rotation curves of low redshift galaxies exhibit a large range of
shapes, and in spite of a large number of attempts \citep[e.g.,][]{persic91},
there is no single expression that is adequate to describe the velocity
gradients of rotationally-supported galaxies with a ``unified'' rotation
curve.  This certainly adds to the uncertainty in interpreting the observed
velocity gradients in high redshift galaxies as rotation, in particular when
the gradient may only be apparently smooth as a consequence of low spatial
resolution.

\citet{nesvadba06} find strong evidence for rotation within the central kpc
(and likely out to radii of a few kpc) of a strongly lensed LBG at z=3.2 --
the ``arc$+$core'' galaxy. In particular, they find that the velocity profile
in the central kpc of the ``arc$+$core'' galaxy is nearly indistinguishable
from the rotation curve of the low-redshift spiral galaxy NGC4419. Contrary
to the arc$+$core, Q0347-383~C5 has a light profile consistent
with two spatially unresolved knots, with an abrupt velocity change
inbetween, which is not suggestive of an isolated rotating disk on the 
scales of a few kpc that we spatially resolve. 

\subsection{The merger scenario}
Based on the emission line morphology and kinematics of Q0347-383~C5,
we favor the merger scenario. With a separation of $\sim 5$ kpc,
the two knots in the emission line image may well be the sites of
two giant star-forming regions, perhaps nuclear starbursts in two
interacting or merging galaxies. Alternatively, the knots may represent
two subclumps formed in a massive, collapsing gas disk, as discussed by
\citet{immeli04} and \citet{bournaud07}. Depending on the initial mass
and the efficiency of energy dissipation of the cold component, the gas
is more or less likely to fragment due to instabilities within an
underlying disk. \citet{bournaud07} argue that this may be a mechanism to
explain 
the irregular, clumpy morphologies of chain and clump-cluster galaxies.
These disk instability models imply disks with numerous small clumps with 10s
of parsecs in size and small masses, to only several but individually large,
1-2 kpc, more massive ($\sim 10^{8-9}$ M$_{\odot}$)
clumps \citep[][]{immeli04,bournaud07}.

Distinguishing between a galaxy merger and the merger of two massive
subclumps embedded within a rotating disk is difficult, but we can give
a tentative answer from our measurements of the velocity dispersion,
estimates of the mass, and upper limits to the sizes of the individual
components within Q0347-383~C5.

LBGs have dynamical masses of on average $\sim 10^{10}$ M$_{\odot}$
\citep{pettini01,nesvadba06}.  Given that the properties of Q0347-383~C5 are
rather typical of other LBGs and that the two components have relatively
similar velocity dispersions, we suspect that our estimated upper limits to
the masses are also within a factor of a few of their actual masses.  Observed
line widths in the integrated spectrum of the lensed ``arc$+$core'' on
physical scales of $\sim 200$ pc are very similar \citep[$\sigma=97\pm 7$ km
s$^{-1}$][]{nesvadba06}, while those observed in $z\sim 2$ galaxies by
\citet{law07,nmfs06} are in many cases significantly larger.  This suggests
that the line widths we observe are representative for the overall widths in
the two knots, and are not an artifact due to the blended kinematics of
neighboring, self-gravitating clouds.

Given the similarities between each component of Q0347-383~C5 and
other LBGs, we favor the interpretation that Q0347-383~C5 consists of
two individual galaxies. Each of these clumps is likely higher in
mass, about a factor of 10 than expected for clumps, and certainly
each has a higher velocity dispersion, $\sim$90 km s$^{-1}$, compared
to 20-30 km s$^{-1}$ for the most massive clumps predicted in the
models simulating disk instability \citep{immeli04}. The small
projected distance of $d_{proj} \sim 5$ kpc and small velocity offset
of $\sim 33\pm 10$ km s$^{-1}$, smaller than the velocity dispersion
of each component, make it unlikely that they lie in a disk
configuration, again suggesting that it is most likely that
Q0347-383~C5 represents a pair of LBGs that will probably merge within
the next few 100 Myrs.

Interestingly, the rest-frame UV absorption lines \citep{steidel96b}
are redshifted relative to the rest-frame optical lines, whereas \lya\
is redshifted relative to the interstellar absorption lines. This may
indicate relatively complex kinematics, with some of the gas infalling
into the system.  Again, this could be taken as evidence in support
of the merger hypothesis, albeit not unique, since an obvious way
of generating gas with a range of ionization and kinematics, including
infall, is through a merger.

\subsection{Is Q0347-383~C5 typical for the luminous tail of the LBG
  population?} 
Is the merger interpretation in agreement with what can be expected from the
global properties of z$\sim3$ LBGs? We address this question by comparing the
number of observed LBGs with properties similar to Q0347-383~C5 with
predictions from hierarchical structure formation models. Recently,
\citet{maller06} predicted the rates of major mergers as a function of
redshift and galaxy mass out to redshifts z$\sim 3$, based on
smoothed-particle hydrodynamical simulations. For baryonic masses greater than
$\sim 2\times 10^{10}$ M$_{\odot}$ they predict a merger rate of $\sim
2-8\times 10^{-4} Gyr^{-1} Mpc^{-3}$, which is less than or roughly similar to
the co-moving space density of LBGs brighter than R$\sim 24$ mag \citep[$\sim
7\times 10^{-4}$ Mpc$^{-1}$][]{adelberger00,reddy07}.  \citet{shapley01} and
\citet{papovich01}, from population synthesis models estimate typical ages of
LBGs at z$\sim$3 to be a few 100 Myrs.  This is roughly half the cosmic time
spanned by the LBG selection which is about 600 Myrs \citep[for $z\sim
2.7-3.4$;][]{steidel03}.  Based on these numbers, \citet{shapley01} estimated
that the ``duty cycle'' for LBGs at z$\sim$3 is about 0.5.

For the merger hypothesis to be viable, mergers must have a similar timescale
during which they could provide the characteristic age and duty cycle observed
in LBGs.  Interestingly, merger timescales are roughly similar to the ages of
LBGs derived from population synthesis models \citep[few$\times 10^8$ yrs,
e.g.,][]{barnes96} and thus we might expect to see about 1/2 the high redshift
population at this epoch undergoing mergers if all bright LBGs have a merger
phase.  

This similarity between timescales and duty cycle suggests that
mergers may play a significant role in the ensemble properties of the LBG
population, in particular within the luminous tail of LBGs, brighter than
$R\sim 24$ mag, and similar to Q0347-383~C5.  \citet{adelberger05b} find for
UV selected galaxies at z$\sim 1.8-3.5$, predominantly at z $\le$ 2.6, that
brighter (at K-band) and redder galaxies have larger correlation lengths than
the fainter ones, suggesting that the more luminous and redder galaxies may
reside in more massive dark-matter halos. Following the hierarchical model,
this would imply that they are presumably more massive and older than their
less massive analogs. Since we did not detect the K-band continuum for
Q0347-383~C5, we cannot compare directly with the \citeauthor{adelberger05b}
results, but suspect that Q0347-383~C5 is significantly fainter than the
K$=$20.5 used by \citeauthor{adelberger05b} to discriminate between faint and
bright UV selected galaxies. 

The irregularity of UV morphologies led \citet{conselice03} to suspect that
major mergers may play an important role at these redshifts, similar to what
we find for Q0347-383~C5. It appears that bright UV selected galaxies have
larger numbers of UV-bright components than their fainter counterparts but
that other morphological and star-formation properties (like the overall
star-formation rate) do not \citep{law07a, shapley03}. While the underlying
processes responsible for these trends are not unambiguously known, at least
these results do not directly contradict the merger hypothesis.

However, \citet{law07a} point out that the absence of clear correlations
between UV morphology and other parameters makes it difficult to associate a
complex continuum morphology in the rest-frame UV uniquely with a merger. They
suggest merger-triggered star-formation should lead to enhanced bolometric
luminosity as well as UV emission from young stellar populations, which they
do not find. However, these studies include ``classical'' LBGs like
Q0347-383~C5, but also galaxies selected with other UV-based
criteria. Overall, this illustrates the difficulties related to purely
morphological and photometric studies and highlights the need to include
integral-field kinematics for statistically robust samples of the various
high-redshift galaxy populations, if we want to understand the underlying
mechanisms governing galaxy evolution in the early universe.

\section{Summary and conclusions}
\label{sec:summary}
We presented an analysis of rest-frame optical integral-field spectroscopy
of the $z=3.23$ Lyman-Break Galaxy Q0347-383~C5 in the K band. This
galaxy is one of the largest known LBGs, and in particular large enough
for seeing-limited observations. Q0347-383~C5 was first described by
\citet{pettini01}, who obtained F702W HST continuum imaging and longslit
spectroscopy in the K-band

We detect the [OIII]$\lambda\lambda$4959,5007 doublet with line
properties that are similar to those discussed in \citet{pettini01}, but
in addition, we also identify H$\beta$ with a flux of $9\times10^{-18}$
erg s$^{-1}$ cm$^{-2}$.  The [OIII]/H$\beta$ line ratio is high, but not
too high for a low-metalicity star-forming galaxy, and corresponds to
an oxygen abundance within the range of metallicities of LBGs measured
by \citet{pettini01}. The observations do not suggest that the optical
spectrum of Q0347-383~C5 is dominated by an AGN.

The [OIII]$\lambda$5007 line image shows two knots at a projected
distance $\sim 0.7$\arcsec\ (5.4 kpc) with a small relative velocity of
33 km s$^{-1}$. Line morphology and kinematics do not resemble those
expected for an outflow or a rotating disk, and more likely originate
from a merger of either two intermediate-mass galaxies with a dynamical
mass of $\le 10^{10} M_{\odot}$ each, or perhaps massive sub-clumps of a
fragmented disk as postulated by \citet{immeli04,bournaud07}. The large
masses of individual knots make it more likely that we see the merging
of two galaxies each tracing its individual dark matter halo or subhalo,
although this is a very difficult distinction to make with present day
data. 
The density of similarly luminous z$\sim 3$ LBGs is consistent
with predictions from recent models of the cosmic evolution of the
merger rate. Star-formation rates estimated from the observed H$\beta$
flux correspond to $\sim 20-40$ M$_{\odot}$ in each clump, which is not
unusual for LBGs generally.

Most z$\sim 3$ LBGs are significantly more compact than Q0347-388~C5,
with typical half-light radii of r$_e\sim 0.3$\arcsec. Such scales are
difficult to resolve with 10-m class telescopes, even with adaptive
optics assisted observations. From such observations
\citet{law07} find that DSF2237a-C2, their only target at z$>$3, has a
velocity gradient and velocity dispersions of the same magnitude as the
shear.  While superficially these characteristics could be suggestive
of a rotating disk, \citet{law07}, from a comparison of their data to
a simple exponential rotating disk model, emphasize that this source is
unlikely to be a thin, rotationally-supported disk. 
Both galaxies are among the largest LBGs and are
comparably bright, which sheds doubts as to whether the properties
of the overall population of $z\sim 3$ LBGs are well described by the
properties of its largest members. \citet{nesvadba06} found evidence for
rotation on sub-kpc scales in a strongly-lensed LBG at z$=3.24$, but
such scales are well beyond reach for generic LBGs even with adaptive
optics. While adaptive optics-assisted observations allow to probe
the dynamics of high-redshift galaxies at sub-kpc resolution, they
must concentrate on galaxies with particularly bright line emission,
to ensure reasonable observing times as pointed out by \citet{law07}.

This will inevitably lead to biases between observed LBG samples and the
parent population of LBGs, and is a reason why studies of gravitationally
lensed are not superceded, but are rather complemented, by high angular
resolution observations of LBGs with adaptive optics, in spite of
uncertainties related to the gravitational magnification.  More positively,
observing galaxies with bright line emission will plausibly provide
information about particularly rapid phases of star-formation and 
galaxy growth, whatever mechanism is responsible for initiating such phases.
Prudence and caution however are certainly justified when generalizing the
results of high redshift galaxies given the current limitation in astronomical
instrumentation and the small sample sizes with detailed 3-dimensional
spectroscopy observations.

\acknowledgements
We would like to thank an anonymous referee for helpful advice and
suggestions that substantially improved this paper and the staff at
Paranal for their help and support in obtaining these observations.
NPHN wishes to acknowledge financial support from the European Commission
through a Marie Curie Postdoctoral Fellowship and MDL wishes to thank
the Centre Nationale de la Recherche Scientifique for its continuing
support of his research.

\bibliography{ms_ref2}

\end{document}